\documentclass[showpacs,10pt,twocolumn,prl]{revtex4-1}
\usepackage{amsmath}
\usepackage{amssymb}
\usepackage{graphics}
\usepackage{epsfig}
\usepackage{color}
\usepackage{amsmath}

\begin{document}

\title{Nematic-fluctuation-mediated superconductivity in Cu$_x$TiSe$_2$}
\author{Xingyu Lv$^{1,2,\dag}$, Yang Fu$^{3,4,1,2,\dag}$, Shangjie Tian$^{5,1,2}$, Ying Ma$^{6}$, Shouguo Wang$^{5}$, Cedomir Petrovic$^{3,4}$, Xiao Zhang$^{6,*}$, and Hechang Lei$^{1,2,*}$}
\affiliation{$^{1}$School of Physics and Beijing Key Laboratory of Optoelectronic Functional Materials \& MicroNano Devices, Renmin University of China, Beijing 100872, China\\
$^{2}$Key Laboratory of Quantum State Construction and Manipulation (Ministry of Education), Renmin University of China, Beijing 100872, China\\
$^{3}$Center for High Pressure Science and Technology Advanced Research, Beijing 100193, China\\
$^{4}$Shanghai Key Laboratory of Material Frontiers Research in Extreme Environments (MFree), Shanghai Advanced Research in Physical Sciences (SHARPS), Shanghai 201203, China\\
$^{5}$Anhui Key Laboratory of Magnetic Functional Materials and Devices, School of Materials Science and Engineering, Anhui University, Hefei 230601, China\\
$^{6}$State Key Laboratory of Information Photonics and Optical Communications \& School of Physical Science and Technology, Beijing University of Posts and Telecommunications, Beijing 100876, China}
\date{\today}
		
\begin{abstract}

The interplay among electronic nematicity, charge density wave, and superconductivity in correlated electronic systems has induced extensive research interest. 			
Here, we discover the existence of nematic fluctuations in TiSe$_2$ single crystal and investigate its evolution with Cu intercalation. 
It is observed that the elastoresistivity coefficient $m_{E_g}$ exhibits a divergent temperature dependence following a Curie-Weiss law at high temperature. 
Upon Cu intercalation, the characteristic temperature $T^{*}$ of nematic fluctuation is progressively suppressed and becomes near zero when the superconductivity is optimized. Further intercalation of Cu leads to the sign change of $T^{*}$ and the suppression of superconductivity.
These results strongly indicate that nematic phase transition may play a vital role in enhancing superconductivity in Cu$_x$TiSe$_2$.
Therefore, Cu$_x$TiSe$_{2}$ provides a unique material platform to explore the nematic-fluctuation-mediated superconductivity.

\end{abstract}
		
%\pacs{74.25.-q, 74.70.Ad, 74.62.Bf}
\maketitle

Electronic nematicity--characterized by spontaneous rotational symmetry breaking of electronic degrees of freedom--emerges in diverse quantum materials. 
For example, cuprate systems such as YBa$_2$Cu$_3$O$_{7-\delta}$ exhibit electronic  nematicity within the pseudogap regime, competing with charge order \cite{daou2010,sato2017,hinkov2008,cyr-choiniere2015}. 
In Sr$_3$Ru$_2$O$_7$, field-induced nematicity appears near metamagnetic quantum critical points \cite{borzi2007}. 
In iron-based superconductors, the nematic fluctuation appears at high temperature when its characteristic temperature $T^{*}$ is closely related to the structural and magnetic transition temperatures $T_{s}$ and $T_{N}$ because of the coupling between electronic nematicity and crystal lattice distortion.
Crucially, the nematic quantum criticality plays a pivotal role in enhancing the superconducting transition temperature $T_c$, particularly in the optimally doped regime
\cite{kuo2013,kuo2016,chu2010,chu2012,palmstrom2022,bohmer2014,bohmer2022}.
Furthermore, orbital-driven nematic phases have been identified in FeSe \cite{baek2015,watson2015}.
Such vital role of nematic fluctuations in enhancing superconductivity has also been validated in kagome metal Cs(V$_{1-x}$Ti$_{x}$)$_3$Sb$_5$ \cite{sur2023}.
Electronic nematicity plays a dual role in superconducting systems: static nematic order generally competes with superconductivity while nematic fluctuations favor it \cite{fernandes2014, lederer2015}.
Thus, exploring  the emergence or enhancement of superconductivity induced by nematic fluctuations or phase transition in other systems is  pivotal for verifying above scenario.

The transition metal dichalcogenide TiSe$_{2}$ represents a prototypical correlated electronic system hosting a commensurate 2 $\times$ 2 $\times$ 2 charge density wave (CDW) order below $T_{\rm CDW} \sim $ 200 K \cite{rossnagel2011,disalvo1976}.
Upon Cu intercalation TiSe$_2$ (Cu$_{x}$TiSe$_{2}$), the CDW is progressively suppressed and a dome-shaped superconducting region with maximum $T_{c} =$ 4 K at $x \sim$ 0.08 appears \cite{morosan2006,wu2007}. 	
Recent theoretical study reveals that electron doping plays a critical role in determining the CDW symmetry and induce both nematic and stripe CDW states of 1$T$-TiSe$_2$ \cite{munoz-segovia2025}.
This naturally rises to a fundamental question: what is the interplay among emergent nematic phase, CDW, and superconductivity in the Cu$_{x}$TiSe$_{2}$?
Elastoresistivity measurement is a powerful technique to probe such static nematic order or fluctuations \cite{daou2010,chu2012,kuo2016,shapiro2015}. 
Moreover, using different symmetry-breaking strain fields, elastoresistivity can be used to distinguish the electronic nematicity in different symmetry channels \cite{chu2012,chu2010,kuo2016,kuo2013,sur2023}.

In this work, we present a systematic study on elastoresistivity of Cu$_{x}$TiSe$_{2}$ single crystals.
For pristine TiSe$_2$, there is a significant response of elastoresistivity in the $E_{g}$ channel and it shows a divergent behavior when temperature approaching the characteristic temperature $T^{*}$, close to $T_{\rm CDW}$, implying the existence of nematic fluctuation in this material. 
With Cu intercalation, the $T^{*}$ shifts to lower temperatures and tends toward zero near optimal superconducting region, manifesting that the enhancement of superconductivity could originate from the nematic phase transition.

\begin{figure*}[tbp]
	\centerline{\includegraphics[scale=0.26]{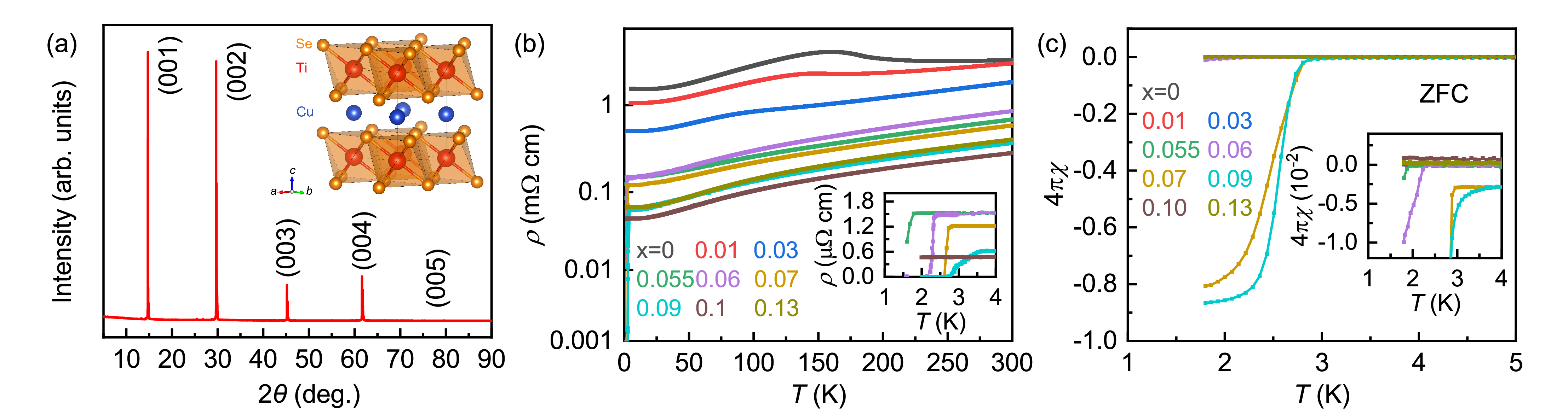}} \vspace*{-0.3cm}
	\caption{
		(a) XRD pattern of a Cu$_{x}$TiSe$_{2}$ single crystal. Inset: crystal structure of Cu$_{x}$TiSe$_{2}$. The small blue, large red and medium orange balls represent Cu, Ti and Se atoms, respectively. 
		(b) Temperature dependence of $\rho_{xx}(T)$ for different Cu content $x$. Inset: enlarged view of $\rho_{xx}(T)$ curves below 4 K.
		(c) Temperature dependence of magnetization $4\pi\chi(T)$ measured at 1 mT with zero-field-cooling (ZFC) mode. Inset: enlarged view of $4\pi\chi(T)$ curves below 4 K.}
\end{figure*}
	
Cu$_{x}$TiSe$_{2}$ single crystals were grown by using chemical vapor transport method. The detailed methods of crystal growth and experimental characterizations are shown in Supplemental Material (SM), Note 1 \cite{SM}.
The crystal structure of Cu$_{x}$TiSe$_{2}$ is shown in inset of Fig. 1(a). It possesses a hexagonal layered structure with space group $P$-3$m$1 (No. 164).  
In each TiSe$_{2}$ layer, the Ti atoms are in octahedral coordination with Se, and  Cu atoms are intercalated in between two TiSe$_{2}$ layers, which are bonded to each other by weak van der Waals interaction. 
Figure 1(a) shows the X-ray diffraction (XRD) pattern of a typical Cu$_{x}$TiSe$_{2}$ single crystal. All of the peaks can be indexed by the indices of (00$l$) lattice planes, indicating that the crystal surface is parallel to the $ab$ plane and perpendicular to the $c$ axis.
Figure 1(b) displays the temperature dependence of the in-plane resistivity $\rho_{xx}(T)$ from 300 K to 1.6 K for Cu$_{x}$TiSe$_{2}$ single crystals with various Cu content. 
For TiSe$_{2}$, there is a broad hump starting from 180 K approximately, related to the CDW transition \cite{morosan2006,wu2007}. 
With increasing Cu content, the CDW transition shifts to lower temperatures and it becomes unobservable for $x\ge$ 0.055.
On the other hand, a superconducting transition appears at low temperature when $x\ge$ 0.055. %coinciding with the disappearance of the CDW. 
At higher Cu content, the superconducting transition temperature $T_{c}$ increases first with maximum value of 3.7 K at $x=$ 0.09, and then decreases gradually. When $x\ge$ 0.1, both CDW and superconductivity can not be observed above 1.6 K, and the samples exhibit a normal metallic behavior. 
The magnetization measurements also provide a consistent evolution of superconductivity with $x$ and the maximum $T_{c}$ is about 3.65 K when  $x=$ 0.09 (Fig. 1(c)).

Because strain $\epsilon$ is a field which can couple to the nematic order parameter $\psi$, the nematic phase transition or fluctuations can be probed by measuring the nematic susceptibility $\chi_{\rm nem} = \partial\psi/\partial\epsilon$ \cite{chu2012, kuo2013}.
Moreover, the $\chi_{\rm nem}$ is proportional to the ratio of resistivity anisotropy to $\epsilon$, i.e., the elastoresistivity coefficients $m_{ij}$ \cite{kuo2013}.
When using Voigt notation, the elastoresistivity coefficients can be expressed as

\begin{equation}
m_{ij} = \frac{\partial(\Delta\rho/\rho)_{i}}{\partial\epsilon_{j}}
\end{equation}

where the indices $i$, $j$  =  1 -- 6 represent the directions $xx$, $yy$, $zz$, $yz$, $zx$ and $xy$, respectively \cite{chu2012}. 
Because the applied $\epsilon$ can be expressed as the sum of several irreducible representations of the crystallographic point group, these $m_{ij}$ can be decomposed into different symmetry channels. 
For $D_{3d}$ point group of Cu$_{x}$TiSe$_{2}$, the $m_{ij}$ associated with the isotropic $A_{1g}$ irreducible representation (irrep) and the anisotropic $E_{g}$ irrep (see detailed discussion in SM, Note 2 \cite{SM}), which are expressed as

\begin{equation}
\begin{aligned}
m_{A1g}&=\frac{(\Delta\rho/\rho_{0})_{1}+(\Delta\rho/\rho_{0})_{2}}{\epsilon_{1}+\epsilon_{2}}\\
&=\frac{(\Delta\rho/\rho_{0})_{1}+(\Delta\rho/\rho_{0})_{2}}{\epsilon_{1}(1-\nu_{p})}\\
&=m_{11}+m_{12}-\frac{2\nu_{s}}{1-\nu_{p}}m_{13}
\end{aligned}
\end{equation}

\begin{equation}
\begin{aligned}
m_{Eg}&=\frac{(\Delta\rho/\rho_{0})_{1}-(\Delta\rho/\rho_{0})_{2}}{\epsilon_{1}-\epsilon_{2}}\\
&=\frac{(\Delta\rho/\rho_{0})_{1}-(\Delta\rho/\rho_{0})_{2}}{\epsilon_{1}(1+\nu_{p})}\\
&=m_{11}-m_{12}
\end{aligned}
\end{equation}

where $\nu_{s}=-\epsilon_{3}/\epsilon_{1}$ and $\nu_{p}=-\epsilon_{2}/\epsilon_{1}$ are the Poisson ratios of  the sample and the piezostacks, respectively.
When existing an electronic nematic fluctuations, the divergence of $\chi_{\rm nem}$ will manifest in a diverging temperature dependence of the $m_{ij}$ in the anisotropic symmetry channel, i.e., $m_{Eg}$.

\begin{figure}[tbp]
	\centerline{\includegraphics[scale=0.32]{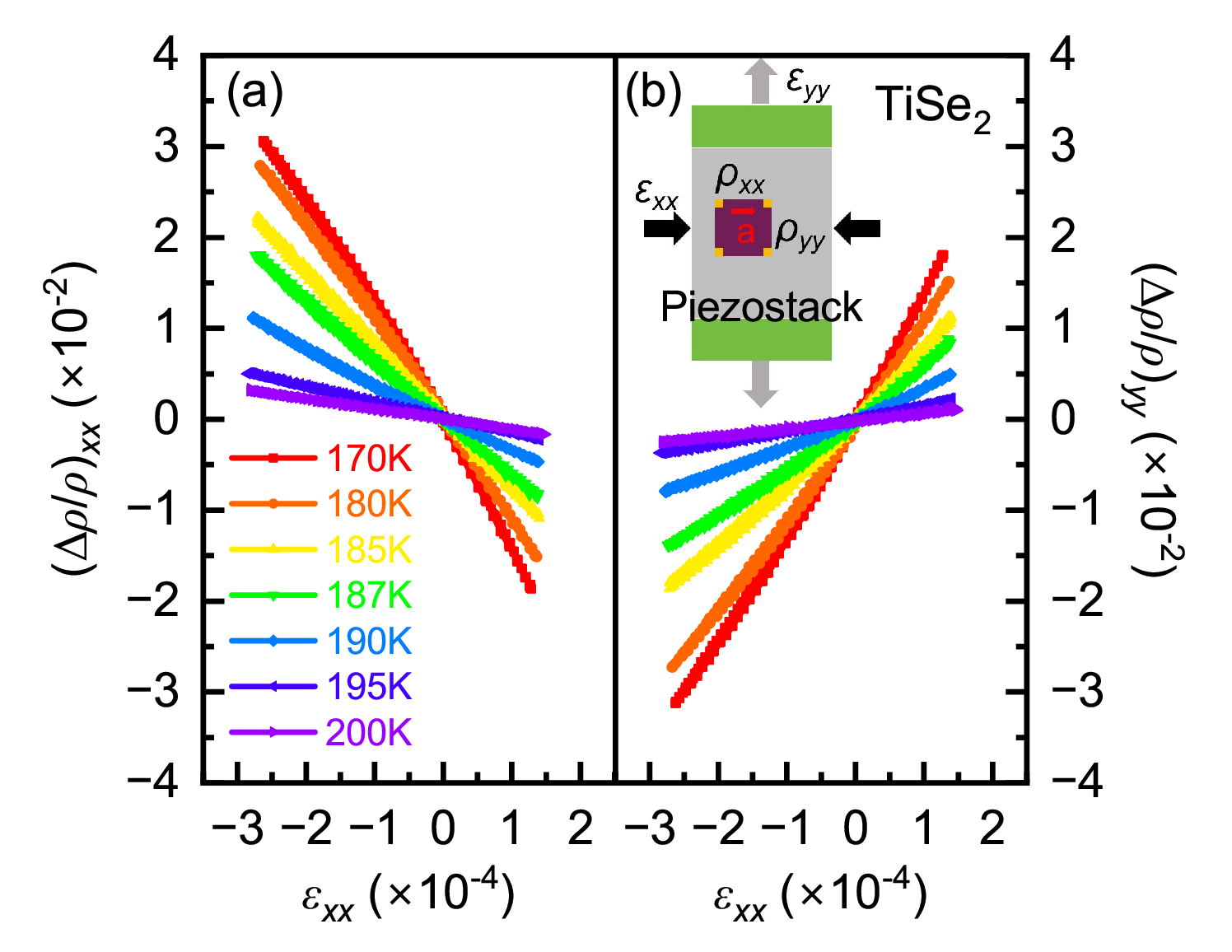}} \vspace*{-0.3cm}
	\caption{
		Relative changes in resistivity (a) ($\Delta\rho/\rho)_{xx}$ and (b) ($\Delta\rho/\rho)_{yy}$ of TiSe$_{2}$ single crystal at various temperatures above $T_{\rm CDW}$ as a function of strain $\epsilon_{xx}$ applied via an attached piezoelectric actuator. 
		Inset of (b) shows a schematic experimental setup using a modified Montgomery technique with the crystallographic $a$-axis parallel to $\epsilon_{xx}$.}
\end{figure}

Figures 2(a) and 2(b) show the representative results of elastoresistivity measurements for pure TiSe$_{2}$ single crystal by using the modified Montgomery method which allows for obtaining $\rho_{xx}$ ($\rho_{1}$) and $\rho_{yy}$ ($\rho_{2}$) from one sample \cite{kuo2016} (see detailed description in SM, Note 3 \cite{SM}). This method does not suffer cross-contamination issues and the symmetry decomposition is exact. 
As shown in the inset of Fig. 2(b), the samples are glued on the sidewall of a piezostack with the crystallographic $a$-axis parallel to $\epsilon_{xx}$ ($\epsilon_{1}$). For this configuration, the purely anisotropic strain $\frac{1}{2}(\epsilon_{1}-\epsilon_{2})$ and isotropic strain  $\frac{1}{2}(\epsilon_{1}+\epsilon_{2})$  can be induced by an external voltage applied to the piezostack. 
It can be seen that  $(\Delta\rho/\rho_{0})_{xx}$ and $(\Delta\rho/\rho_{0})_{yy}$ exhibit a linear dependence of $\epsilon_{xx}$ with rather weak hysteresis. This implies that all of measured elastoresistivity coefficients are in the linear response regime. For the former, the sign of slope is negative when it is positive for the latter.
Importantly,  the absolute values of slopes of $(\Delta\rho/\rho_{0})_{xx}(\epsilon_{xx})$ and $(\Delta\rho/\rho_{0})_{yy}(\epsilon_{xx})$ show a strong temperature dependence, i.e., they increase quickly with decreasing temperature. Similar behaviors have been observed in Cu intercalated TiSe$_{2}$ single crystals (Fig. S1 in SM \cite{SM}).

\begin{figure*}[tbp]\centering%
\centerline{\includegraphics[scale=0.25]{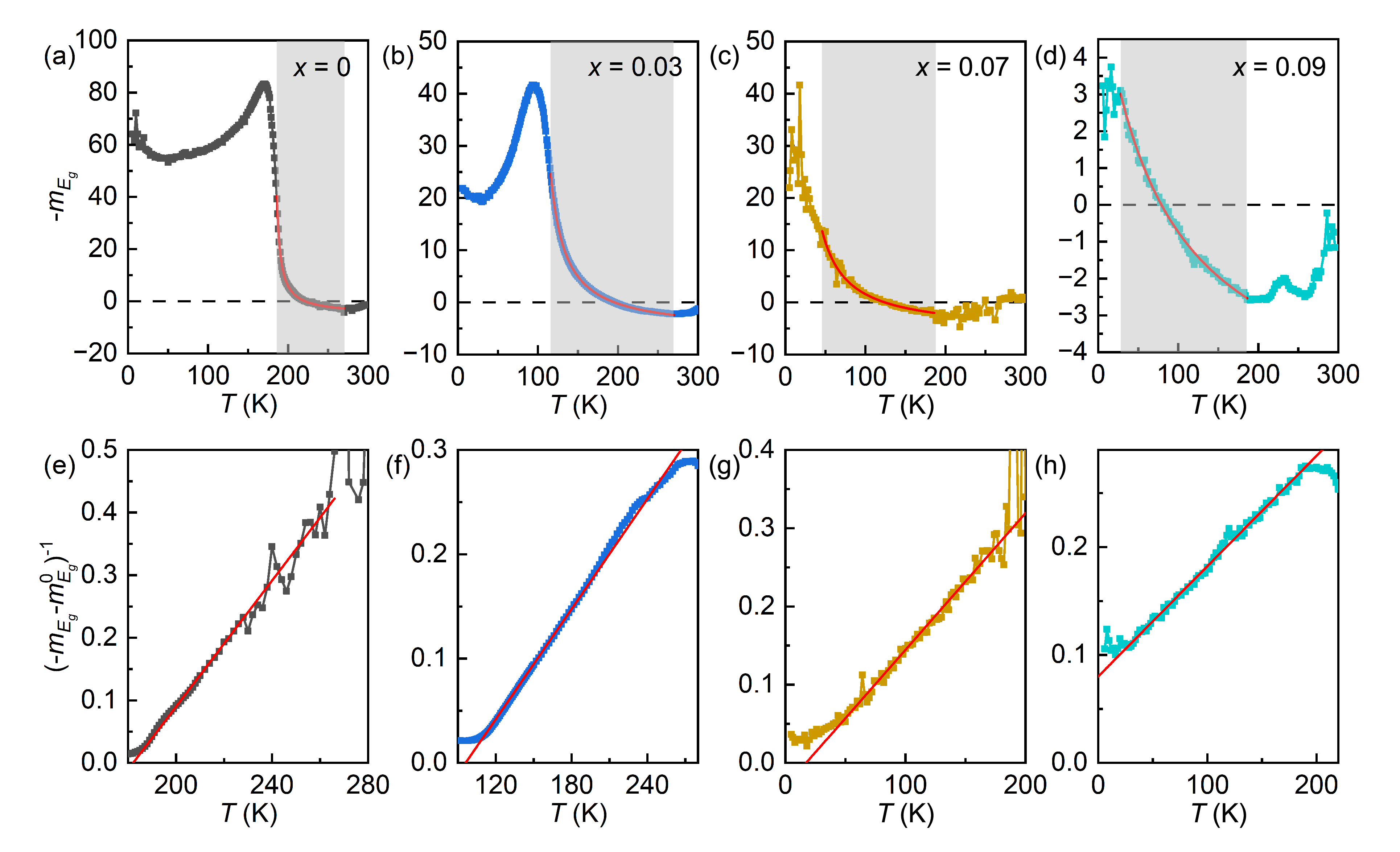}} \vspace*{-0.3cm}
\caption{(a) -- (d) Temperature dependence of $-m_{E_{g}}(T)$ of Cu$_{x}$TiSe$_{2}$ for $x$ = 0, 0.03, 0.07 and 0.09. (e) -- (h) Corresponding $(-m_{Eg}-m_{Eg}^{0})^{-1}$ as a function of temperature. The red solid lines represent the fits using the CW formula and the gray-shaded areas in (a) -- (d) display the fitted regions.}
\end{figure*}

From the linear fits of $(\Delta\rho/\rho_{0})_{xx}(\epsilon_{xx})$ and $(\Delta\rho/\rho_{0})_{yy}(\epsilon_{xx})$ curves, the $m_{A1g}$ and $m_{Eg}$ can be calculated using Eqs. (2) and (3). 
Figure 3(a) and Fig. S2 in SM \cite{SM} show the temperature dependence of $-m_{E_{g}}$ and $m_{A1g}$ of TiSe$_{2}$. 
It can be seen that the coefficient of $-m_{E_{g}}$ increases rapidly when temperature decreases to $T_{\rm {CDW}}$, and then start to decrease with further lowering temperature, i.e., there is a peak in the $-m_{E_{g}}(T)$ curve, the temperature of which is close to the $T_{\rm CDW}$. At $T<$ 50 K, the $-m_{E_{g}}$ increases slightly again.
In contrast, the $m_{A1g}(T)$ shows a much weaker temperature dependence and smaller values than those of $-m_{E_{g}}(T)$. It only exhibits a kink when $T$ is close to $T_{\rm {CDW}}$.
It is presumably related to critical fluctuations, similar to the behavior of $m_{11}-m_{12}$ in Ba(Fe$_{0.975}$Co$_{0.025}$)$_{2}$As$_{2}$ \cite{kuo2013}.
Hence, it could be concluded that there exists the diverging elastoresistivity coefficient in the anisotropic channel $m_{E_{g}}$. To gain more insight, we fit the high-temperature $m_{E_{g}}(T)$ curve of TiSe$_{2}$ using a Curie-Weiss (CW) temperature dependence \cite{chu2012,kuo2013},

\begin{equation}
\begin{aligned}
-m_{Eg}=\dfrac{\lambda}{a(T-T^{*})}+m^{0}_{Eg} 
\end{aligned}
\end{equation}

where $\lambda/a$ is the Curie constant, and $T^{*}$ is the Weiss temperature. 
As shown in Fig. 3(a), the $-m_{E_{g}}(T)$ curve can be fitted perfectly (red line), and the linear temperature dependence of the inverse susceptibility $(-m_{Eg}-m^{0}_{Eg})^{-1}$ further reflects the validity of CW behavior (Fig. 3(e)).
The fitted $T^{*}$ is 181.59(9) K. Similarly, the $-m_{E_{g}}(T)$ as a function of $T$ can be extracted from the elastoresistivity measurements for the Cu-intercalated TiSe$_2$ crystals, which are shown in Figs. 3(b) -- 3(d) and Fig. S3 in SM \cite{SM}. 
The nematic susceptibility curves of Cu$_{x}$TiSe$_{2}$ with $x=$ 0.01 and 0.03 samples closely resemble that of TiSe$_{2}$. But their peaks shift to lower temperatures (135 K for $ x=$ 0.01 and 92 K for $x=$ 0.03).
With higher Cu content ($x\geq$ 0.055), the peak feature cannot be observed in the whole temperature range and the $m_{E_{g}}(T)$ curves exhibits monotonic increases upon cooling, consistent with the behavior of $\rho_{xx}(T)$ curves (Fig. 1(b)) because of the suppression of CDW with Cu intercalation. 
In contrast, the nematic fluctuations persist in the samples with much higher $x$ (= 0.09). 
%Moreover, the temperature regime of these fluctuations systematically enlarges to lower temperature with increasing Cu content. 
It is noted that the absolute values of $m_{E_{g}}(T)$ decrease with increasing $x$ in general and the fluctuation behavior vanishes when $x\geq$ 0.1 (Fig. S3 in SM \cite{SM}).
By using the fits according to eq. (4), the obtained $T^{*}$ decreases monotonically with the increase of $x$ from 155.0(1) K for $x$ = 0.01 to 13(3) K for  $x$ = 0.07. The $T^{*}$ becomes negative when $x$ increases further (-77(5) K for $x=$ 0.09).

\begin{figure}[tbp]
\centerline{\includegraphics[scale=0.3]{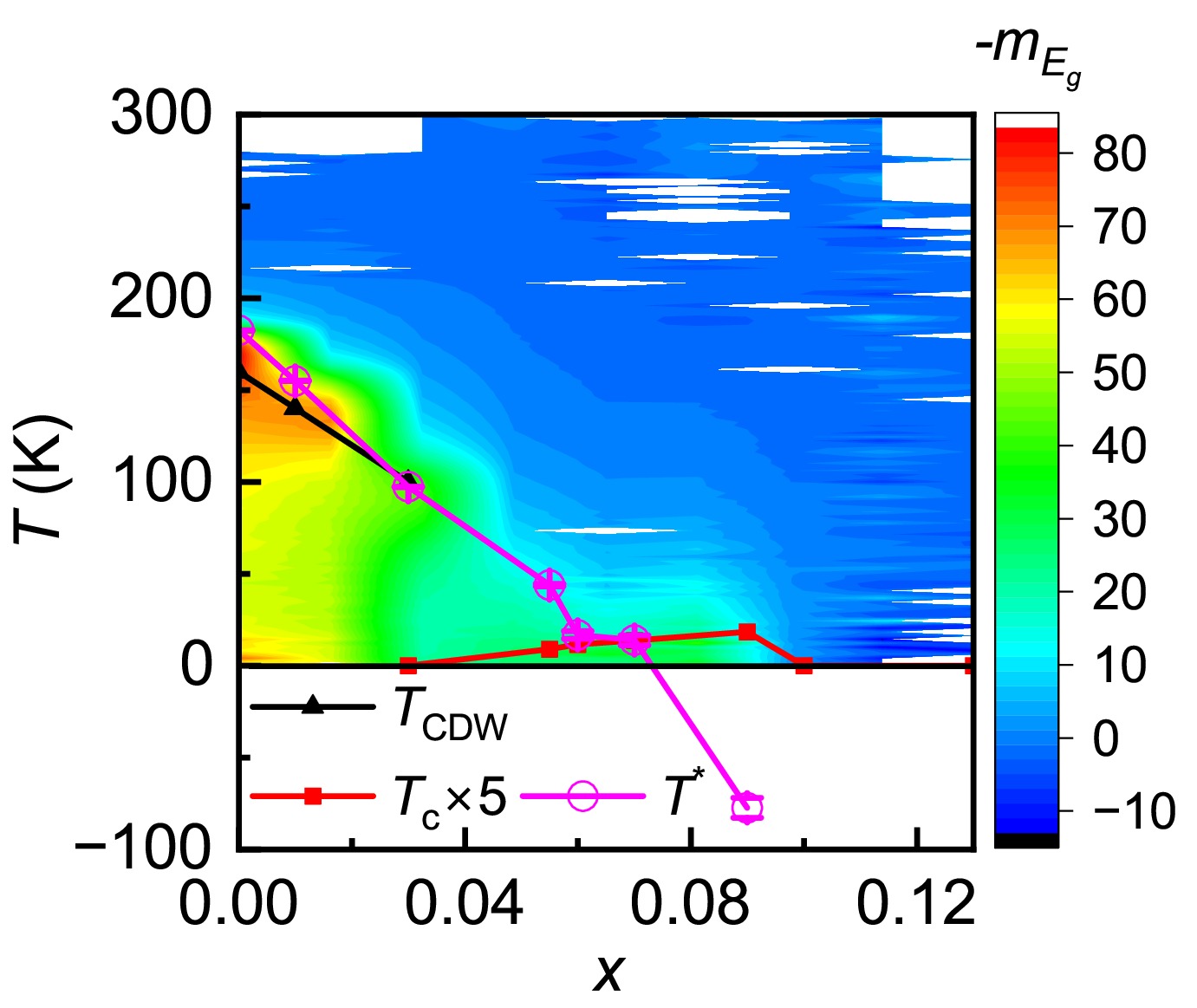}} \vspace*{-0.25cm}
\caption{Phase diagram of Cu$_{x}$TiSe$_{2}$. The $T^{*}$ is denoted by pink open circles (error bars indicate fitting uncertainty). $T_{\rm {CDW}}$ and $T_c$ are represented by black triangles and red squares, respectively. The color scale shows the magnitude of  $-m_{E_{g}}$.}
\end{figure}

To investigate the relationships between nematicity fluctuation, CDW and superconductivity, the evolution of $T^{*}$, $T_{\rm {CDW}}$ and $T_c$ with Cu content is summarized in a phase diagram (Fig. 4).
In the electron underdoped regime, both $T^{*}$ and $T_{\rm CDW}$ decrease with increasing $x$, and the $T^{*}$ closely tracks the $T_{\rm CDW}$ until the CDW state disappears ($x$ = 0.055), where superconductivity appears simultaneously.
With approaching optimal doping level ($x\sim$ 0.08) \cite{morosan2006,wu2007} where the superconductivity of Cu$_{x}$TiSe$_{2}$ becomes bulk and the $T_c$ reaches its maximum value ($\sim$ 4 K), the $T^{*}$ decreases further to near zero.
The $T^{*}$ even becomes negative as the doping further increases beyond optimal doping level, indicating a paranematic state. Similar behaviors have been observed in iron-based superconductors \cite{chu2012,kuo2016}.
These results strongly suggest that the nematic phase transition is intimately related to the enhancement of $T_c$ in Cu$_{x}$TiSe$_{2}$.

In summary, we investigated the evolution of electronic nematicity in Cu$_x$TiSe$_2$ via elastoresistivity measurements. 
It is found that the temperature-dependent $m_{E_g}(T)$ exhibits a CW behavior at high-temperature region, demonstrating the existence of nematic fluctuations in this system.
With Cu intercalation, the characteristic temperature $T^*$ of nematic fluctuations is suppressed to zero and changes its sign near the optimal superconducting region, manifesting the essential role of nematic phase transition in enhancing superconductivity in Cu$_{x}$TiSe$_{2}$.
       
\section{Acknowledgments}

This work was supported by the National Key R\&D Program of China (Grants Nos. 2023YFA1406500 and 2022YFA1403800), the National Natural Science Foundation of China (Grant Nos. 12274459 and 52572288), the Shanghai Key Laboratory of Novel Extreme Condition Materials, China (Grants No. 22dz2260800), and the Shanghai Science and Technology Committee, China (Grants No. 22JC1410300). 

$^{\dag}$ X.Y.L. and Y.F. contributed equally to this work.

$^{\ast}$ Corresponding authors: X.Z. (zhangxiaobupt@bupt.edu.cn); H.C.L. (hlei@ruc.edu.cn).

\end{document}